%% file: document.tex
\documentclass[article]{jss}

\usepackage{thumbpdf,lmodern}

\usepackage{mypackages}

\author{Christoph Lienhard
   \And Torsten A. En{\ss}lin}
\Plainauthor{Christoph Lienhard \And Torsten A. En{\ss}lin}

\title{Hamiltonian Monte Carlo Sampling for Fields}
\Plaintitle{Hamiltonian Monte Carlo Sampling for Fields} 
\Shorttitle{HMC Sampling for Fields}

\Abstract{
  \hmcif\ ``Hamiltonian Monte Carlo for Fields'' is a software add-on for
  the \nifty\ ``Numerical Information Field Theory'' framework
  implementing Hamiltonian Monte Carlo (HMC) sampling in \python.
  \hmcif\ as well as \nifty\ are designed to address inference problems
  in high-dimensional spatially correlated setups such as image reconstruction.
  They are optimized to deal with the typically high number of degrees of freedom.

  \hmcif\ adds an HMC sampler to \nifty\ that
  automatically adjusts the many free parameters
  steering the HMC sampling machinery.
  A wide variety of features ensure efficient
  full-posterior sampling for high-dimensional inference problems.
  These features include integration step size adjustment, evaluation of the
  mass matrix, convergence diagnostics, higher order symplectic integration and
  simultaneous sampling of parameters and hyperparameters in Bayesian
  hierarchical models.
}

\Plainkeywords{Python, field inference, Bayesian statistics, Hamiltonian sampling}

\Address{
  Christoph Lienhard\\
  Max Planck Institut for Astrophysics\\
  Karl-Schwarzschild-Str. 1\\
  D-85741 Garching, Germany\\
  \emph{and}\\
  Ludwig-Maximilians-Universit\"at M\"unchen\\
  Geschwister-Scholl-Platz 1\\
  80539 Munich, Germany\\
  \emph{and}\\
  Technische Universit\"at M\"unchen\\
  Arcisstr. 21\\
  80333 Munich, Germany\\
  E-mail: \email{christoph.lienhard@tum.de}\\

  Torsten En{\ss}lin\\
  Max Planck Institut for Astrophysics\\
  Karl-Schwarzschild-Str. 1\\
  D-85741 Garching, Germany\\
  \emph{and}\\
  Ludwig-Maximilians-Universit\"at M\"unchen\\
  Geschwister-Scholl-Platz 1\\
  80539 Munich, Germany\\
}

\begin{document}

\input{./tex/introduction}
\input{./tex/installation}

\input{./tex/problem_description}
\input{./tex/examples}

\input{./tex/conclusion}


\section*{Acknowledgments}
We would like to thank all members of the Information Field Theory group at MPA
for many useful discussions in particular, and alphabetical order:
Philipp Arras, Jakob Knollm\"uller, Daniel Pumpe, Reimar Leike and Martin Reinecke.

\newpage
\bibliography{refs}


\newpage

\begin{appendix}

\input{./tex/epsilon_adjustment}
\input{./tex/software_structure}

\end{appendix}


\end{document}

%% file: tex/introduction.tex
\section{Introduction}\label{sec:intro}

\subsection{Purpose}
\hmcif\ implements a Hamiltonian Monte Carlo (HMC) sampler \citep{duane1987}
for the \nifty\ (``Numerical Information Field Theory'') framework
\citep{selig2013, steininger2017, nifty4}).
It is available for \pkg{Python3} on Unix-like systems.
The main purpose of \hmcif\ is to create samples which are distributed
according to arbitrary, once-differentiable probability distributions.
These samples can, for example be used to approximate expectation values in
cases where brute-force integration is infeasible.

\nifty\ is a \python\ library developed for computational work with
information field theory (IFT, \citet{ensslin2009, ensslin2010}).
IFT extends classical probability theory onto functional spaces.
\nifty\ is interesting for spatially correlated inference
problems such as image reconstruction \citep{d3po, resolve, d4po} or work on
geospatial datasets \citep{pysesa}.
A main advantage is the resolution-independent calculation of statistical
estimates.

With \hmcif , Bayesian models already implemented in \nifty\ can 
be reused for an HMC sampling approach, easily.
This can help estimating the impact of approximations present in other
approaches, or enable tackling entirely new problems.

\subsection{Features}
At the heart of \hmcif\ lies the \hmcsc\ class which constructs an HMC sampler
based on a predefined \nifty\ \eon\ class describing a probability
distribution $\Prob(x)$ as an \emph{energy} $\Psi(s) = - \log (\Prob(x))$.
Samples drawn from this distribution are saved to the disk as \hdf\
archives \citep{hdf}.

To ensure a successful sampling process \hmcif\ implements a variety of
additional features.
The sampler calculates a convergence measure to determine when the burn-in
phase has finished.
Several predefined strategies on how to exactly calculate the measure are
available and can be chosen from.
It is critical for an HMC sampler to use a proper \emph{mass matrix} which is
why \hmcif\ can recalculate it several times during burn-in achieving better
performance by orders of magnitude in comparison to the usage of an identity as
mass matrix.
Another important sampling parameter, the \emph{integration step size} of the
symplectic integrator, is also adjusted such that a predefined acceptance rate
is matched.
Again, the exact adjusting strategy can be chosen from a predefined set of
options.
Although for most purposes a second order leapfrog integrator is sufficient,
higher order integrators  are available as
well.
Furthermore, \hmcif\ uses multi-processing in that individual Markov chains use
separate cores.

\hmcif\ is optimized to ease the work with HMC sampling.
All of the above can be done in only a few lines of code
if a well-defined \nifty\ \eon\ class is available.

\subsection{Comparison to other Packages}
There are many software packages for HMC sampling available in many different languages.
But unlike \hmcif\ most packages are static in that they use in general the identity
as the mass matrix or need a mass matrix specified in the beginning.
Since especially in high-dimensional cases a good mass matrix estimation
is crucial for a successful sampling process we will concentrate
on packages which estimate the mass matrix.

A very popular cross-platform package for HMC is \stan\ \citep{stan}.
\stan\ provides interfaces for \proglang{R}, \python , \proglang{shell},
\proglang{MATLAB}, \proglang{Julia}, \proglang{Stata}, \proglang{Mathematica}
and \proglang{Scala}.
Its biggest advantage is the \proglang{C++} back-end which makes it by far
the fastest sampler if the same parameters such as integration step size
and mass matrix are chosen.
Another notable advantage over \hmcif\ is an implementation of the no-u-turn
sampler (NUTS, \citet{hoffman2014})
which can be seen as an extension to the standard HMC approach.

In \stan\ the mass matrix is set to the identity initially, but is recalculated
from the generated samples during the burn-in phase.
The mass matrix can but does not have to be restricted to a diagonal matrix.
\hmcif\ differs in that the user is able to define their own
mass matrix which can be an advantage in some cases (see e.g. \citet{taylor2008}).
The \stan\ developers announced such a feature in future versions, though.
Using the samples generated by the initial chain itself involves the risk of having
highly correlated samples in case the sampler was malfunctioning due to the
usage of the initial mass matrix.
To avoid this, \hmcif\ uses samples drawn from the curvature of the full posterior at
the current position to reevaluate the mass matrix.
We found this approach to be much more efficient.
Reevaluated mass matrices are always diagonal in \hmcif\ but since it is targeted at
high-dimensional problems where non-sparse matrices can not be represented explicitly
this is not really a disadvantage.
Furthermore, more recent \nifty\ based algorithms use 
harmonic space degrees of freedom as field variables \citep{knollmuller2017}
which fits better to a mass matrix being diagonal in these field parameters.

Another important package for HMC sampling in \python\ is \pymc\ \citep{pymc}.
\pymc\ provides a huge variety of different samplers among other functions for
statistical applications.
When it comes to the HMC sampler in \pymc\ the main difference to \hmcif\
is that the mass matrix is again evaluated based on the samples of the Markov
chain itself which might be problematic as described in the paragraph above.
Again \pymc\ has a NUTS feature.

Apart from that the main advantage of \hmcif\ is that it is easy to use for
already written algorithms in \nifty\ and its optimization for high-dimensional
statistical problems.

\subsection{Structure of this Paper}
This introduction is followed by a short installation guide.
In section \ref{sec:theory} we give an introduction to HMC sampling on a
theoretical / mathematical level.
Afterwards, we illustrate the work with \nifty\ and \hmcif\ using a simple
Bayesian hierarchical model as an example in section \ref{sec:eg}.
This document ends with a short summary in section \ref{sec:summary} on
why there is a need for a distinct \hmcif\ package. 

%% file: tex/installation.tex
\section{Installation}\label{sec:install}

\subsection{Dependencies}
\hmcif\ relies on the following other \python\ packages:
\begin{description}
\item[\np ]: The very basic \python\ package for numerical analysis
  on multidimensional arrays.
\item[\pkg{SciPy}] \citep{oliphant2007} : Library implementing advanced
  algorithms and functions in \python .
\item[\pkg{h5py}]: A \python\ wrapper for the HDF5 file format.
\item[\nifty] (4.1.0 or newer, \citet{steininger2017, nifty4}) :
  A package for statistical field inference problems.
\item[\pkg{matplotlib}] \citep{hunter2007}, optional :
  A package for producing figures.
  Necessary for the \hmcif\ \pkg{tools} sub-package.
\item[\pkg{PyQt5}] optional : Necessary for the \pkg{tools} sub-package.
\end{description}

\nifty\ supports multi-processing in many calculations via \pkg{mpi4py}
(\cite{mpi4py}) but \hmcif\ needs to restrict
each individual Markov chain to one core.
If \pkg{mpi4py} is already installed
the user should switch multi-processing off by setting
the environment variables \code{MKL_NUM_THREADS} and
\code{OMP_NUM_THREADS} to 1 in a terminal:
\begin{CodeChunk}
  \begin{CodeInput}
  export MKL_NUM_THREADS = 1
  export OMP_NUM_THREADS = 1
  \end{CodeInput}
\end{CodeChunk}

\subsection{Installing via Git}
Installing \hmcif\ along with all required packages is possible via
\begin{CodeChunk}
  \begin{CodeInput}
  pip install git+https://gitlab.mpcdf.mpg.de/ift/HMCF
  \end{CodeInput}
\end{CodeChunk}

%% file: tex/problem_description.tex
\section{Hamiltonian Monte Carlo Sampling}\label{sec:theory}
The goal of most statistical problems is to find expectation values
$\left\langle f(x) \right\rangle_{\Prob(x)}$ of a function $f(x)$
given a distribution $\Prob (x)$, with
\begin{equation*}
  \left\langle f(x) \right\rangle_{\Prob(x)} = \int_\paramspace f(x) \Prob(x) dx \,,
\end{equation*}
where $\paramspace$ is the space of all possible values for $x$.
However, especially in high dimensional cases the integral may become
intractable.
One approach to circumvent this problem is to use a form of Monte Carlo
integration.
Samples $(x_i)$ which are distributed like $\Prob(x)$ are used
to approximate the expectation value:
\begin{equation}\label{eq:approxexpval}
  \left\langle f(x) \right\rangle_{\Prob(x)} \approx \frac 1N \sum_{i=1}^N f(x_i)
\end{equation}
The law of large numbers ensures that this approximation converges to the true
expectation value in non-pathological situations.
Then, the problem is reduced to finding a strategy to generate the samples
$(x_i)$.
While there are straight forward solutions in some cases such as normal
distributions, often more advanced algorithms are needed.
A large subset of such algorithms are Markov chain Monte Carlo (MCMC) methods
which have in common that a Markov process is constructed which ultimately
converges to the wanted \emph{target distribution} $\Prob(x)$.
HMC sampling is a MCMC method especially applicable for once-differentiable
probability distributions on high-dimensional spaces.

\subsection{The Algorithm}\label{sec:hmcalgo}
The Hamilton Monte Carlo (HMC) approach (first introduced by \citet{duane1987},
good summaries: \citet{neal2011, betancourt2017})
uses a variation of the Metropolis-Hastings algorithm \citep{hastings1970,
metropolis1953} with less random walk behavior.
The idea is to describe the Markov process as a physical Hamiltonian time
evolution, exploiting for MCMC methods advantageous properties of the dynamics
of this system.
The samples $(x_i)$ can then be thought of as snapshots of the trajectory of a
particle moving through an energy landscape $\Psi(x)$.
This \emph{energy} is defined by the target distribution $\Prob(x)$ via
\begin{equation*}
 \Psi(x) := - \log(\Prob(x))
\end{equation*}
This convention originates from statistical physics where the probability of a
system being in state $i$ with energy $E(i)$ is
\begin{equation*}
  \Prob(x) \propto \exp(-\beta E(i))
\end{equation*}
where $\beta$ is a temperature-dependent scaling parameter.

Additionally, a new normal distributed random variable $p \in \paramspace$ with
covariance $M$ called \emph{momentum} is introduced.
The negative logarithm of the joint distribution $\Prob (x, p)$ then looks like
a typical physical Hamiltonian:
\begin{equation*}
 H(x, p) = \frac 12 p^\top M^{-1} p + \Psi(x) + \text{const}
\end{equation*}
The central idea of HMC sampling is to evolve this system in time
according to Hamilton's equations of motion
\begin{equation}\label{eq:hamiltondyn}
  \begin{split}
 \dot x^k &= ~~ \frac{\partial H}{\partial p^k} ~ =  \left[  M^{-1}p\right]^k\\
 \dot p^k &= - \frac{\partial H}{\partial x^k} ~ = - \frac{\partial \Psi}{\partial x^k}
  \end{split}
\end{equation}
for $k = 1, \ldots, \text{dim}(\paramspace )$.
After some time $T$ the position we arrived at is considered as a new
Metropolis-Hastings proposal $(x(T), p(T)) =: (x^\prime, p^\prime)$.
This is approach is possible since Hamiltonian dynamics have some convenient
properties such as volume preservation.
Also, in theory, the new sample is exactly as probable as the starting point
since the process is energy conserving.
In practice however, the discretization of the integration in time leads to
errors in the energy conservation, which is why a accept-reject step is necessary
where the proposal is accepted with probability
\begin{equation}\label{eq:accprobhmc}
  \rho_A = \min \left( 1, \exp (-\Delta E) \right) \,,
\end{equation}
where $\Delta E = H(x^\prime, p^\prime) - H(x, p)$.

The whole algorithm then looks like this:

\begin{algorithm}[H]
Set initial $x_{0}$\\
\For{$i=1$ \KwTo \#samples}{
  $\left.
    \begin{tabular}{lll}
      Generate momentum sample $p \sim \mathcal N (0, M)$\\
      Evolve system for time $T$\\
      New position: $(x^\prime, p^\prime) = (x(T), p(T))$\\
    \end{tabular}
  \right\}$ MH proposal\\
  \vspace{0.5em}
  Generate sample $r \sim \mathcal U ([0,1])$\\
  \eIf{$r \leq \rho_A$}{
    Set $x_{i} = x^\prime$
    }{
    Set $x_{i} = x_{i-1}$
    }
}
\end{algorithm}
The resampling of $p$ in each iteration ensures that the whole parameter space
$\paramspace$ can be reached.
At this point we omit the full proof that the samples $(x_i)$ are then
distributed like $\Prob (x)$. See \citet{neal2011} for details.

\subsection{Further Technicalities}
For HMC to work, the integrator for the time evolution of the system needs to be
symplectic.
Most of the time the \emph{leapfrog} integrator is used, which is a second order
symplectic integrator.
Symplectic integrators of higher orders based on work presented in
\citet{higherorder} are possible as well in \hmcif .
They have proven to be advantageous for HMC sampling in high-dimensional
non-linear cases \citep{blanes2014}.

One step in time of length $\epsilon$ with the leapfrog integrator is calculated
via
\begin{equation}\label{eq:leapfrog}
\begin{split}
 p\left(t+\frac \epsilon 2\right) &= p(t) - \frac \epsilon 2 \left. \frac{\partial \Psi}{\partial x} \right|_{x(t)} \\
 x\left(t+ \epsilon \right) &= x(t) + \epsilon M ^{-1}p \\
 p\left(t+ \epsilon \right) &= p\left(t + \frac \epsilon 2\right) - \left. \frac\epsilon 2 \frac{\partial \Psi}{\partial x} \right|_{x(t+\epsilon)}\,.
\end{split}
\end{equation}
This single leapfrog step is applied $L$ times to generate a new sample such
that $T = \epsilon L$.
The integration step size $\epsilon$ determines the overall acceptance rate of
the sampling process.
The advantages of HMC are only present if the acceptance rate is in the range of
0.6 to 0.9 \citep{accrange01, accrange02}.
To relate $\epsilon$ to the acceptance rate we developed an approximation
further discussed in appendix \ref{sec:epsadj}.

Finally, for an HMC sampling process to work properly it is crucial to find a
good mass matrix $M$, which serves as covariance of the momentum $p$.
There are several approaches but one very popular strategy is
to use samples from the chain itself and base the mass matrix on the variance of
these samples.
\begin{equation}\label{eq:masseval}
  M^{-1} = \frac 1N \sum_{i=1}^N \left( (x_i-\mu)(x_i-\mu)^\top \right)
\end{equation}
with $\mu$ being the mean value.
However, in specific cases other approaches might be better,
for example as documented in \citet{taylor2008}.
We found that using samples from the chain itself is unfeasible in
high-dimensional cases ($\text{dim}(\paramspace)>10000$) because a bad initial
mass matrix leads to extremely small integration step sizes and highly
correlated samples.
Thus, in \hmcif\ the samples are generated by drawing samples from the curvature
of $\Psi(x)$ at the current position of the chain.
In other words: For the purpose of estimating the mass matrix, we approximate
the target distribution $\Prob (x)$ with a normal distribution and then draw
samples from this distribution.

%% file: tex/examples.tex
\section[Using \purehmcif ]{Using \hmcif }\label{sec:eg}

\input{tex/demo_multi_field}

%% file: tex/demo_multi_field.tex
In this section we try to provide some intuition in working with \hmcif .
It is assumed that the reader is familiar with \python .

We will use a simple Bayesian hierarchical model as an example to introduce the
most important functionalities of \hmcif .
For a comprehensive documentation of all features see appendix \ref{sec:soft}.
An implementation of this example is part of the \hmcif\ repository and can be
found in \code{demos/multi_field_demo.py}.
We first introduce the model on a theoretical level, then give a short overview
on how the model is implemented within \nifty\ in section \ref{sec:mfimpl}.
In section \ref{sec:hmcifsampling} we show how sampling with \hmcif\ works and
finally introduce features for retrieving and displaying the results in section
\ref{sec:hmcifresults}.

\subsection{A Simple Hierarchical Model}
Consider a telescope observing the large-scale structure of the universe.
Billions of galaxies producing photons eventually reaching the earth.
This photon flux reaching the sensors of the telescope is denoted $x$.
The spatial properties of $x$ can be described as a mathematical field living on
a continuous space.

Our telescope measures this photon flux $x$ which means that it converts it into
an electronic signal based on how many photons reach each of its CCD sensors.
Errors in lenses and small differences between individual sensors alter the true
nature of the original $x$ field but can be coped with by calibration.
In our model, this transformation is represented by a linear \emph{response}
operator $R$ acting on $x$.
Note, that the domain and the target domain of $R$ are different.
Whereas, in theory, the domain of $x$ is continuous, the output of the telescope
is definitely discretized.
Additionally to the response of the instrument, we assume a certain Gaussian
noise due to imperfect electronics with known covariance $N$.
The measured data $d$ produced by the telescope is then described by the
measurement equation
\begin{equation}\label{eq:wienermodel}
  d = R(x) + n\,.
\end{equation}

What we are interested in is the ``true'' signal, based on the data $d$ we got
from the telescope.
This information is reflected in the conditional probability $\Prob(x | d)$.
In terms of Bayesian statistics this is often referred to as the
\emph{posterior} and Bayes formula relates the posterior to our assumed model
and prior knowledge we may have on our signal $x$:
\begin{equation*}
  \Prob(x | d) \propto \Prob(d | x) \Prob(x)
\end{equation*}

While the \emph{likelyhood} $\Prob(d | x)$ is defined by our model in equation
(\ref{eq:wienermodel}) and the noise statistics, the \emph{prior} $\Prob(x)$
needs more consideration.
Observations dating back to \citet{hubble1934} suggest that for the large-scale
structure, at least to some extend, a log-normal prior is sufficient.
We therefore introduce another field $s = \log (x)$ with a multivariate Gaussian
distribution and covariance $S$ such that $x$ is log-normal distributed.
We further want to ensure that our field $s$ is somewhat smooth.
This can be enforced by imposing a power law $P(k)$ on the covariance $S$ in its
harmonic representation,
\begin{equation*}
  S_{kk^\prime} = \delta_{k,k^\prime} P(k)\,.
\end{equation*}
This power law can be chosen such that high curvatures in the position space get
punished and are therefore improbable.
For illustration, we assume a power law
\begin{equation}\label{eq:mfpowlaw}
  P(k) = \left( \frac{\lc}{1 + \lc k} \right)^4
\end{equation}
with the \emph{correlation length} $\lc$ essentially defining how strong
curvature is allowed to be.

If we do not know the correlation length, we can treat it as a free hyperparameter
making the problem a full Bayesian hierarchical model.
Since $\lc$ is strictly positive we assume another log-normal prior
\begin{equation*}
  \Prob(\lc) \propto \frac 1 {\lc} \exp \left( - \frac {(\log \lc - \mu_c)^2}{2 \sigma^2_c}\right)
\end{equation*}
where $\mu_c$ and $\sigma^2_c$ are appropriate parameters for this hyperprior.

Our full posterior then turns into
\begin{equation}\label{eq:mfpost}
  \Prob(s, \lc | d) \propto \Prob(d | s)\Prob(s| \lc) \Prob(\lc )
\end{equation}
with the likelihood
\begin{equation*}
  \begin{split}
  \Prob(d | s) &= \int \delta (d - Re^s - n) \Prob(n)  dn\\
  &\propto \exp \left( - \frac 12 \left( d - Re^s \right)^\top N^{-1} \left( d - Re^s \right)\right)
  \end{split}
\end{equation*}
and the prior for $s$
\begin{equation*}
  \Prob(s | \lc) = \left| 2 \pi S \right|^{-\frac 12} \exp \left( - \frac 12 s^\top S^{-1} s \right)
\end{equation*}
The dependence on $\lc$ is encoded in the covariance $S$.
The norm of $S$ is defined as the determinant
\begin{equation*}
  |S| = \det S = \prod_k P(k)
\end{equation*}

However, for HMC sampling we need a potential $\Psi(s, \lc)$ i.e., the negative
logarithm of the posterior in (\ref{eq:mfpost}).
For better readability, we divide the potential in a prior and a likelihood part
as well, such that
\begin{equation}\label{eq:mfpot}
    \Psi(s, \lc) = \Psil (s, \lc) + \Psip(\lc)
\end{equation}
with
\begin{equation*}
  \begin{split}
    \Psil(s, \lc) &= - \log \Prob (d | s) - \log \Prob(s | \lc)\\
    \Psip(\lc) &= - \log \Prob(\lc)
  \end{split}
\end{equation*}
We omit terms constant in $\lc$ and $s$ since they are not important for HMC
sampling and arrive at 
\begin{equation}\label{eq:mfenergies}
  \begin{split}
    \Psil(s, \lc) &= \frac 12 \left( d - Re^s \right)^\top N^{-1} \left( d - Re^s \right)
      + \frac 12 \left( s^\top S^{-1}s + \sum_k \log (P(k)) \right)\\
    \Psip(\lc) &= \frac 12 \left(  \frac 1{\sigma^2_c}(\log \lc - \mu_c)^2\right) + \log(\lc)
  \end{split}
\end{equation}

Additionally, the gradient of the potential $\Psi (s, \lc)$ is needed for the
time evolution part during HMC sampling.
For the likelihood part the gradient boils down to the following expressions:
\begin{equation}\label{eq:mflikegrad}
  \begin{split}
    \partial_s \Psil &=  S^{-1}s - \left( R e^s \right) \odot N^{-1}\left( d - Re^s \right) \\
    \partial_{\lc}\Psil &= s^\top \left(\partial_{\lc}S^{-1}\right)s + 4 \sum_k \left( \frac 1{\lc} - \frac{k}{1 + \lc k} \right)
  \end{split}
\end{equation}
where $\odot$ denotes point-wise multiplication of vectors.
For deriving the exact expression for $\partial_{\lc}S^{-1}$, observe that
$\left[ S^{-1} \right]_{kk^\prime} = \delta_{kk^\prime} \left(  P(k)
\right)^{-1}$ and therefore
\begin{equation*}
  \left[ \partial_{\lc}S^{-1}\right]_{kk^\prime} = -4 \frac{(1+\lc k)^3}{\lc^5} \delta_{kk^\prime}
\end{equation*}
For the prior part of the potential the gradient can be written as
\begin{equation}\label{eg:mfpriorgrad}
  \begin{split}
    \partial_s \Psip &= 0 \\
    \partial_{\lc}\Psip &= \frac 1{\lc}\left(\frac 1{\sigma_c^2} ( \log \lc - \mu_c ) -1 \right)\,.
  \end{split}
\end{equation}

\subsection[Implementation within NIFTy]{Implementation in \nifty }
\label{sec:mfimpl}
In this section we will take a look at how such a model can be implemented in
\nifty\ in general.
For more details on \nifty\ see \citet{nifty4}.

In \nifty\ there are a variety of classes representing certain aspects of a
typical inference problem, among which the most important ones are:
\begin{description}
\item [\code{Domain} :] A base class representing the underlying space.
  For example a regular grid can be defined as a \code{RGSpace} which inherits
  from \code{Domain}.
\item [\code{Field} / \code{MultiField} :] A class representing fields such as
  $x$, $n$ or $d$.
  It carries information about the underlying \code{Domain} as well as the
  field's value on this domain.
  The \code{Field} supports every arithmetic operation.
  Other operations such as trigonometric or exponential functions can be applied
  point-wise using e.g.,  \code{ift.exp(x)}.
  The notation for these functions is the same as the one used by \np .
  \code{MultiField}s are essentially dictionaries of \code{Field}s which also
  support all operations above.
  This can be used to represent all free parameters in a hierarchical model in
  just one object.
\item [\code{LinearOperator} :] An abstract base class for explicitly or
  implicitly defined linear operators such as the response $R$.
  The \code{LinearOperator} class carries information about the operator's
  \code{Domain} as well as its target domain.
  The operator can be applied to a \code{Field} \code{x} in various ways by
  calling one of the methods \code{times(x)}, \code{inverse_times(x)},
  \code{adjoint_times(x)}, or \code{adjoint_inverse_times(x)}, although not
  every of these methods is available for every linear operator.
\item [\eon\ :] An abstract base class representing the negative
  logarithm of a distribution $\Prob(x)$.
  The \eon\ class is only defined at a certain value of a \code{Field} or
  \code{MultiField} but the same energy for a different \code{Field y} on the
  same \code{Domain} can be obtained by calling the \code{at(y)} method.
  Additionally, the \eon\ class defines a gradient at the position as well as
  the curvature.
  This class is the most important one for \hmcif\ since it defines the
  potential $\Psi(s, \lc)$ and thereby the whole model.
\end{description}

The model introduced in the previous section can be implemented as an \nifty\
\eon\ but since this paper is about \hmcif\ we will not go into detail with
this.
The curious reader can, however, look into the demonstration script and will
find the implementations of the likelihood and the prior energy from
(\ref{eq:mfenergies}) along with their respective gradients in
(\ref{eq:mflikegrad}) and (\ref{eg:mfpriorgrad}) in \code{/demos/energies.py}
in the package's repository.

\subsubsection[Implementation of the Hierarchical Model in NIFTy]
{Implementation of the Hierarchical Model in \nifty }
This and the following section summarize the \code{demos/multi_field_demo.py}
script.
The reader may want to have a look at the script, to have an overview.

We first import \nifty\ and \hmcif\ among other packages via
\begin{CodeChunk}
  \begin{CodeInput}
import nifty4 as ift
import hmcf
  \end{CodeInput}
\end{CodeChunk}

\begin{figure}[h]
  \centering
  \includegraphics[clip, trim=4.5cm 0cm 3.5cm 0cm]{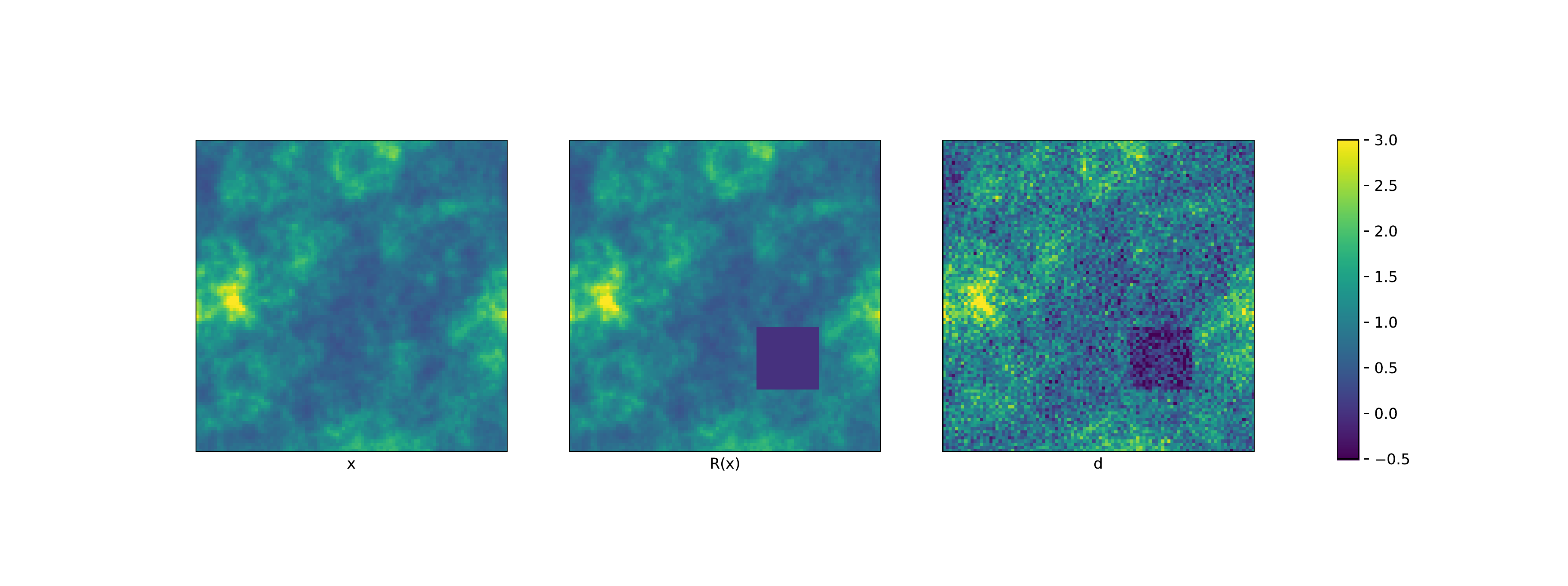}
  \caption{Output of the \code{get\_hierarchical\_data} function.
    The first image on the left displays the original photon flux $x$ before it
    hits the detector.
    For this simple example the instrument transforms the photon flux perfectly
    to an electrical signal, except for a square in the bottom left region of the
    image where it is broken.
    The instrument returns just zeros in that region.
    The resulting image is displayed in the middle figure.
    Finally in the right figure, the data field $d$ is displayed where Gaussian
    random noise was added.
    }
  \label{fig:mfsource}
\end{figure}

A mock dataset for our algorithm to operate on is generated using the
\code{get_hierarchical_data} function which was written for this purpose only.
\begin{CodeChunk}
\begin{CodeInput}
d, sl, R, N = get_hierarchical_data()
\end{CodeInput}
\end{CodeChunk}
It returns \nifty\ objects for the data field $d$, the free parameters $s$ and
$\lc$ (as the \code{MultiField sl}), the response $R$ and the noise covariance
$N$.
\code d is a \code{Field} containing the data, \code{sl} is a \code{MultiField}
with entries \code{'signal'} for $s$ and \code{'l_c'} for $\lc$.
The mock dataset is generated by first sampling a signal
field $s$ with the power spectrum defined in (\ref{eq:mfpowlaw}) and a
pre-defined value for $\lc$.
Together with a noise drawn from a normal distribution with covariance $N$ the
measurement equation (\ref{eq:wienermodel}) is applied and the mock data set is
available.
The signal space as well as the data space consist of $100\times 100$ pixels
which means that, together with $\lc$, our problem has $10,001$ free parameters.
In figure \ref{fig:mfsource} this mock data set generated with random seed $41$
can be observed.
For this simple example the instrument $R$ transforms the photon flux perfectly
to an electrical signal, except for a square in the bottom right region of the
image where the instrument is broken and just returns zeros.
For our sampling process we wrote an \niftye\ class called
\code{SimpleHierarchicalEnergy} which implements the potential $\Psi(s, \lc)$
described in (\ref{eq:mfpot}).
The constructor needs the following arguments:
\begin{description}
  \item[\code{position} :] 
    \nifty\ \code{MultiField}\\
    The position $(s, \lc)$ where the energy $\Psi(s, \lc)$ is evaluated.
    The \code{MultiField} has two entries: \code{'signal'} and \code{'l_c'}.
    The \niftyf\ in \code{'signal'} is the harmonic representation of the $s$
    parameter.
  \item[\code{d} :]
    \niftyf \\
    The data vector.
  \item[\code{R} :]
    \niftyo \\
    The instrument's response.
  \item[\code{N} :]
    \niftyo \\
    The covariance of the noise.
  \item[\code{l\_c\_mu} :]
    \float \\
    Hyperparameter for the log-normal distribution of $\lc$.
  \item[\code{l\_c\_sig2} :]
    \float \\
    Hyperparameter for the log-normal distribution of $\lc$.
  \item[\code{HT} :]
    \niftyo \\
    Harmonic transformation operator, capable of transforming the
    \code{position['signal']} field from the harmonic representation to the
    position space.
  \item[\code{inverter} :]
    \nifty\ \code{Minimizer} \\
    Numerical method for inverting \niftyo s.
    This is necessary to be able to sample from the curvature of the energy
    which is required if the mass matrix is supposed to be reevaluated.
\end{description}
Using a \code{MultiField} as position enables us to sample the signal and
the hyperparameter $\lc$ at the same time.
\code{d}, \code{R} and \code{N} are already given by the
\code{get\_hierarchical\_data} function.
Values for \code{l\_c\_mu} and \code{l\_c\_sig2} are chosen such that values for
$\lc$ in the range of $0.05$ to $10.$ are probable (the true value used during
data generation is $0.6$).
A harmonic transformation operator is easily defined via
\begin{CodeChunk}
\begin{CodeInput}
s_space = sl['signal'].domain
HT = ift.HarmonicTransformOperator(s_space)
\end{CodeInput}
\end{CodeChunk}
and as the inverter needed for the mass reevaluation we use a conjugate gradient
implementation in \nifty :
\begin{CodeChunk}
\begin{CodeInput}
ICI = ift.GradientNormController(iteration_limit=2000,
                                 tol_abs_gradnorm=1e-3)
inverter = ift.ConjugateGradient(controller=ICI)
\end{CodeInput}
\end{CodeChunk}
Finally, for the initial definition of the energy we use the \code{sl} as
position since it has the correct \code{MultiField} structure.
An instance of the energy for our model is then created by calling the
constructor
\begin{CodeChunk}
\begin{CodeInput}
energy = SimpleHierarchicalEnergy(sl, d, R, N, HT, -0.3, 2.,
                                  inverter=inverter)
\end{CodeInput}
\end{CodeChunk}

\subsection[Sampling with \purehmcif ]{Sampling with \hmcif}
\label{sec:hmcifsampling}
Up to this point we only introduced and used \nifty\ objects.
But now that an \eon\ is properly defined we can start using \hmcif .
First, we create an instance of the \hmcsc\ class.
This object represents the HMC sampler and the constructor has only one required
argument:
The potential $\Psi$.
However, for this example we also set the optional \code{num\_processes}
argument which states the number of chains or CPU cores we use during sampling
and the \code{sample\_transform} argument which is necessary since we sample $s$
but are mainly interested in the photon flux $x = \exp(s)$.
\begin{CodeChunk}
\begin{CodeInput}
def sample_trafo(s):
    val = dict(s)
    val['signal'] = ift.exp(HT(val['signal']))
    return ift.MultiField(val=val)

sampler = hmcf.HMCSampler(energy, num_processes=6, 
                          sample_transform=sample_trafo)
\end{CodeInput}
\end{CodeChunk}

The \code{sample_transform} argument requires a function and represents
essentially $f$ in (\ref{eq:approxexpval}).
It is important that the \code{sample_transform} function takes \niftyf s or
\code{MultiField}s of the same kind as the position argument of the \eon\
instance which is passed to the constructor of the \hmcsc\ class (in particular
they need to live on the same domain).
The output of the function can be any kind of \code{Field} or \code{MultiField}
regardless of what was put in.

Before we start sampling we need to define initial positions for the Markov
chains.
In principle this could be completely random but unfortunately, in
high-dimensional cases we need to be somewhere close to non-vanishing parts of
the target distribution because otherwise the gradients are so large that
numerical integration during time evolution will fail in any case.
In this example we know the true signal \code{sl} and will use small derivations
from that, but under real circumstances one would need to first use an
approximating algorithm to get close to the mean or maximum-a-posteriori
solution of the problem and start from there.
This can be done using algorithms already implemented in \nifty .
Afterwards, we call the \code{run} method of our sampler which has again only one
required argument: The number of samples per Markov chain drawn after the
burn-in phase has finished.
Additionally, we set the optional argument \code{x_initial} with a list of
initial positions (the length of this list does not have to be the same as the
number of Markov chains).
\begin{CodeChunk}
\begin{CodeInput}
x_initial = [sl*c for c in [.5, .7, 1.5, 2.]]
sampler.run(500, x_initial=x_initial)
\end{CodeInput}
\end{CodeChunk}
This will initiate a sampling process where the sampler starts in a burn-in
phase during which the integration step size $\epsilon$ from (\ref{eq:leapfrog})
is adjusted to meet the default target acceptance rate of $0.8$.
Additionally, the mass matrix is reevaluated once in the beginning and then the
sampler waits until the individual Markov chains have converged with respect to
a measure based on diagnostics first introduced by \cite{gelman1992}.
All these things can be adjusted to the users needs and specific problem and a
detailed description of all options can be found in appendix \ref{sec:soft}.
The whole sampling process may take up to 10 minutes depending on the machine
the script is executed on.
If the sampling process seems to freeze in the beginning this is probable due to
the mass reevaluation which can take some time.
A much shorter execution time is possible by setting the \code{n\_pixels}
argument of the \code{get\_hierarchical\_data} function to \code{10}.

\subsection{Retrieving Results after Sampling}\label{sec:hmcifresults}
After some time the sampler is finished and one can access the mean value (of
the transformed samples) as a \nifty\ \code{MultiField} via the \code{mean}
attribute of the \code{sampler}.
To get the photon flux values as a \np\ array one can write
\begin{CodeChunk}
\begin{CodeInput}
mean_val = sampler.mean['signal'].to_global_data()
\end{CodeInput}
\end{CodeChunk}
The \code{'signal'} statement selects the \niftyf\ representing the signal $s$
and the \code{to_global_data()} method returns the \code{Field}'s value on the
regular grid as a \np\ array.
To get the inferred mean value of the correlation length $\lc$ as a \float\ the
following statement does the trick:
\begin{CodeChunk}
\begin{CodeInput}
l_c_val = sampler.mean['l_c'].to_global_data()[0]
print(l_c_val)
\end{CodeInput}
\begin{CodeOutput}
0.637494646985
\end{CodeOutput}
\end{CodeChunk}
The same thing is possible by loading the mean value from the respective \hdf\
file:
\begin{CodeChunk}
\begin{CodeInput}
mean_val = hmcf.load_mean(path/to/file)['signal']
l_c_val = hmcf.load_mean(path/to/file)['l_c'][0]
\end{CodeInput}
\end{CodeChunk}
Obviously, this is possible even if the \code{HMCSampler} instance was deleted
already (for example after a reboot of your system).

\begin{figure}[ht]
\centering
\includegraphics[clip, trim=2cm 0cm 1.5cm 0cm]{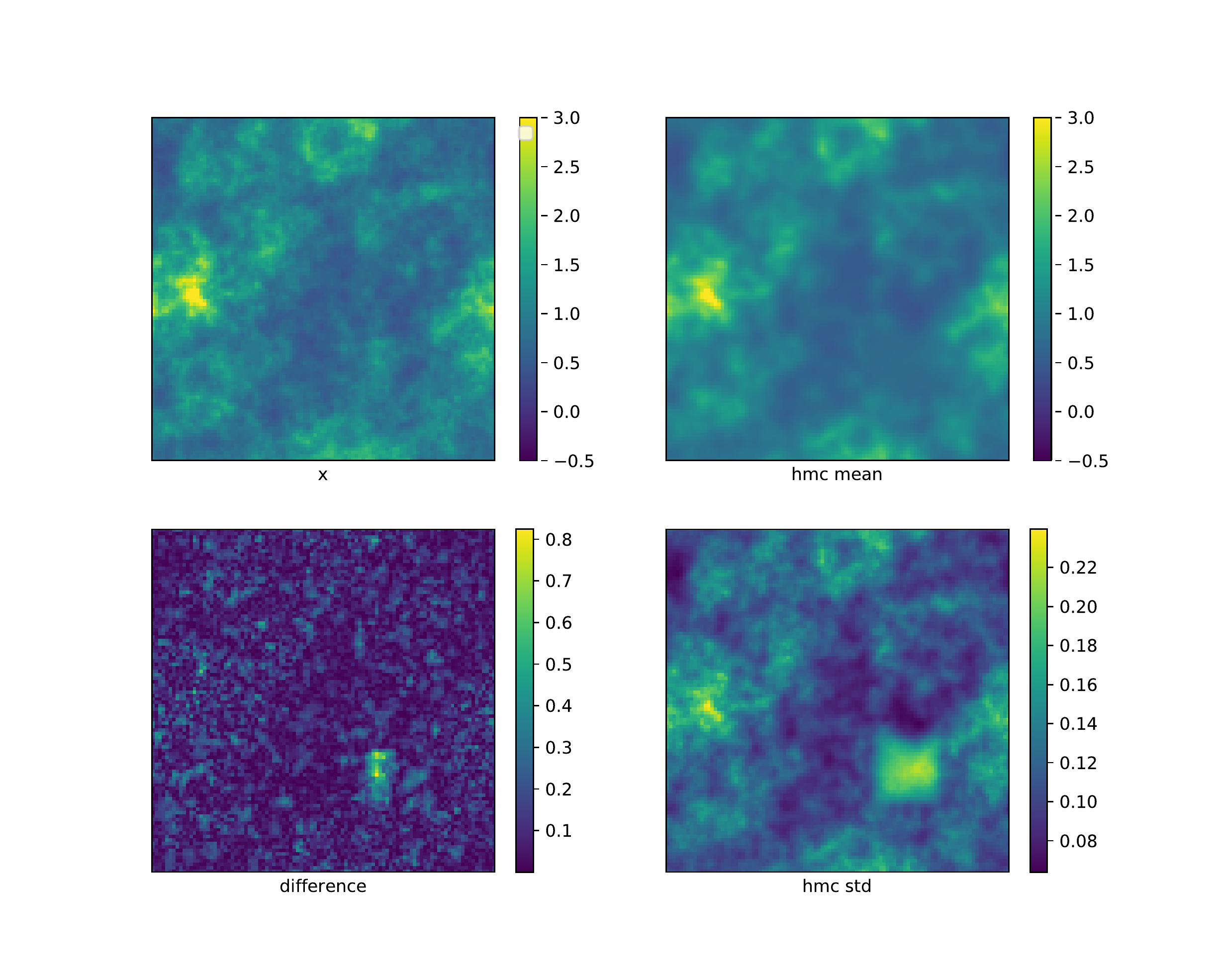} 
\caption{Results of the multi\_field\_demo.py script.
  In the upper row the true photon flux $x$ along with the reconstructed picture
  based on the samples generated by the HMC algorithm.
  The reconstructed picture gets very blurry where the instrument was broken in
  the bottom right region and information got lost.
  In the second row the absolute difference between the original flux and the
  reconstructed picture and the standard deviation of the HMC samples is
  displayed.
  Again, the region in which the instrument is broken is very prominent in both
  cases.}
\label{fig:mfres}
\end{figure}

The variance of the samples can be obtained in the same way using the \code{var}
attribute of the \hmcsc\ class or calling the \code{load_var} function.
The results of the sampling process are displayed in figure \ref{fig:mfres}.

The most prominent difference between the original flux $x$ and the HMC mean
value is where the instrument was broken in the bottom right region.
In particular, the standard deviation of the samples drawn from the full
posterior distribution is remarkably high there as one would expect since
information is missing.

The samples itself can be loaded by either using the \code{samples} attribute of
the \hmcsc\ class or calling the \code{load} function of \hmcif .
As an example the shape of the returned \np\ array can be displayed with
\begin{CodeChunk}
\begin{CodeInput}
print(hmcf.load(path/to/file.h5)['signal'].shape)
\end{CodeInput}
\begin{CodeOutput}
(6, 500, 100, 100)
\end{CodeOutput}
\end{CodeChunk}
where the first element reflects the number of independent Markov chains, the
second element is the number of sample each chain generated and the third and
fourth element reflect the regular grid on which the whole process was carried
out.

\begin{figure}[ht]
  \centering
  \includegraphics[width=0.9\textwidth]{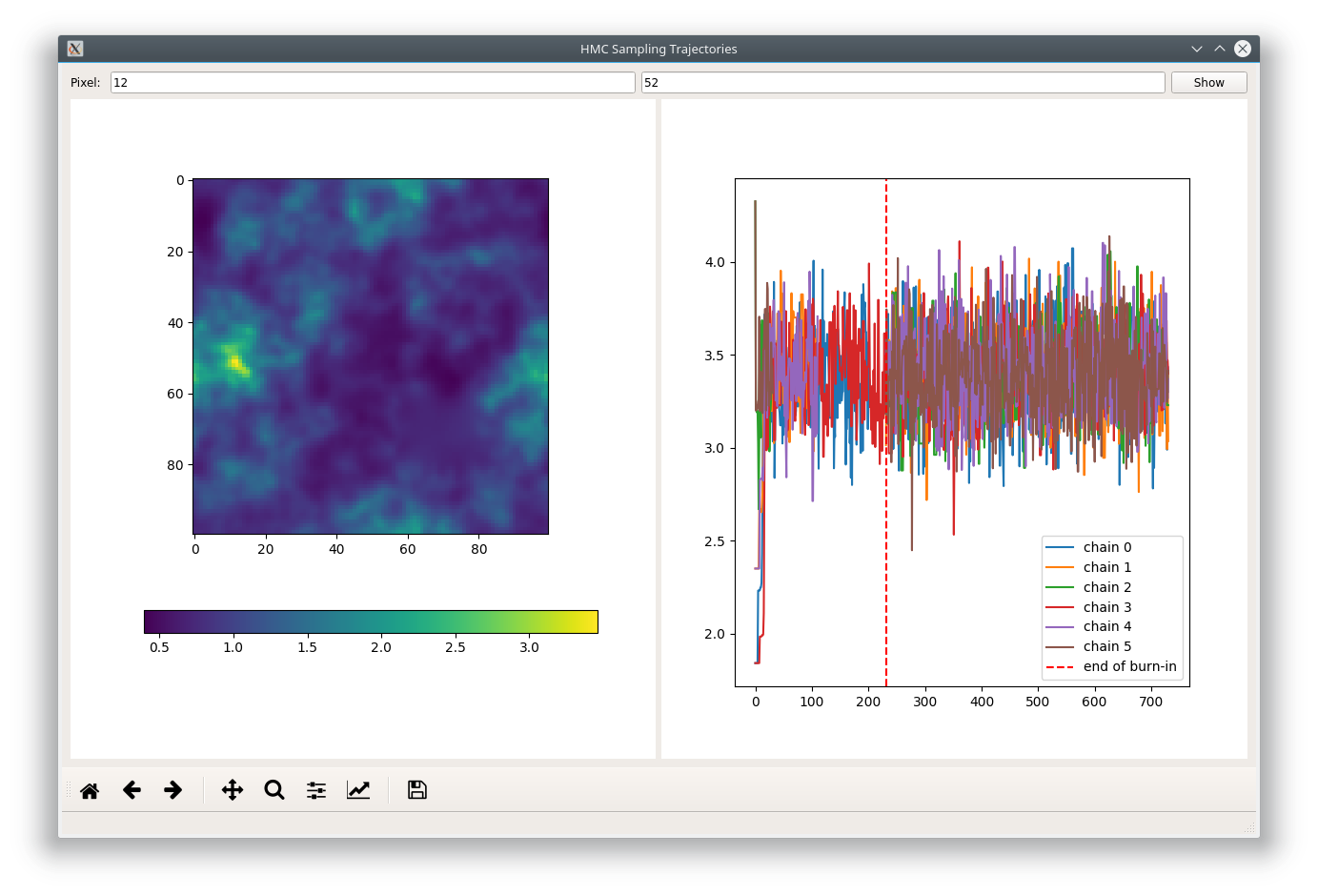} 
  \caption{The evaluation GUI for displaying the Markov chain trajectories
    of selected pixels. On the left the mean value for $s$ in our example is
    displayed.
    In the right graph the trajectories of the six Markov chains at the pixel
    coordinates $(x=12, y=52)$ (as stated in the top row) are shown.}
  \label{fig:gui01}
\end{figure}

To have a better overview of these samples we can use the
\code{show\_trajectories} function of \hmcif\ which displays the chain
trajectories through parameter space by pixel.
It is an interactive GUI consisting of two graphs as depicted in figure
\ref{fig:gui01}.
The left graph shows the inferred mean value and the right graph trajectories of
each chain for one selected pixel.
The pixel can either be set in the top row by defining the coordinates and then
clicking on ``show'' or by just clicking on some pixel in the left graph.
The function itself is located in the \hmcif\ \code{tools} sub-package and can
be called by executing
\begin{CodeChunk}
\begin{CodeInput}
from hmcf.tools import show_trajectories

show_trajectories(path/to/file, field_name='signal')  
\end{CodeInput}
\end{CodeChunk}
where the \code{field\_name} statement defines which element of the
\code{MultiField} is displayed.

%% file: tex/conclusion.tex
\newpage 
\section{Summary}\label{sec:summary}
Efficient HMC sampling with the high number of degrees of freedom
of a numerically represented field is a very complicated task.
\hmcif\ takes care of most challenges arising while working on such problems.
It provides good default values and adjusting strategies for crucial
parameters such as the integration step size or the mass matrix.
Nonetheless the user is still able to customize many details
of how the sampler deals with a given problem.
These features include
\begin{itemize}
\item Simultaneous sampling of all free parameters and hyperparameters
\item Setting the order of the symplectic integrator
\item Defining the adjustment strategy for $\epsilon$ and related properties
\item Defining the convergence measure strategy and related properties
\item Defining how often and how well the mass matrix is reevaluated
\item Providing a clear, in-depth overview of relevant parameters of all chains
  especially during burn-in phase
\end{itemize}

Apart from a diverse set of different options to choose from
the structure of the module even eases the creation of new, customized
options.
We explained the usage of \hmcif\ and demonstrated its performance using the
demonstrator coming with the \hmcif\ package.

%% file: tex/epsilon_adjustment.tex
\section{Dynamic Step-Size Adjustment}\label{sec:epsadj}
Hamiltonian Monte Carlo needs much more computation time per sample proposal than other
MCMC approaches.
This is a disadvantage that is compensated by much higher acceptance rates and
lower autocorrelation between samples.
Because of the energy conserving property of Hamiltonian dynamics,
the acceptance rate is only dependent on the performance of the numerical integrator.
The numerical integrator has one free parameter: the step size $\epsilon$.
In principle the bigger it is the bigger is the integration error $\Delta E$
and thereby the smaller the acceptance rate.
On the other hand, a too small value for $\epsilon$ results in a sampler which does not
explore the typical set on bearable timescales.
In practice an acceptance rate of 0.7 to 0.9 \citep{accrange01, accrange02} is
said to be ideal for an HMC sampler to make up for the higher computation time.

To this end we adjust $\epsilon$ during the burn-in phase such that a
user-defined acceptance rate is matched.
The first step in constructing a good strategy is to derive a relation between
acceptance rate $r_A$ and the integration error $\Delta E$.
We developed the following approximation.
Given the acceptance probability $\pacc$ from equation (\ref{eq:accprobhmc}),
the \emph{expected} acceptance rate is given by
\begin{equation}\label{eq:accrate}
  r_A(\epsilon) = \left\langle \pacc (\Delta E) \right\rangle_{P(\Delta E | \epsilon)}
  = \left\langle \min \left( 1, e^{-\Delta E} \right) \right\rangle_{P(\Delta E | \epsilon)}
\end{equation}
where $P(\Delta E | \epsilon)$ is the probability distribution for $\Delta E$
conditioned on $\epsilon$.

To tackle the $\min$ function properly it is assumed that the probability
distribution for the sign of $\Delta E$ is not dependent on the absolute value 
of $\Delta E$:
\begin{equation*}
  P(\Delta E) \approx P(|\Delta E |) P(\sgn(\Delta E))
\end{equation*}
This reflects the plausible situation that errors are symmetrically probable displacements
of trajectories in regions of the phase space that are dominated
by a potential gradient and not by a minimum.
In this case we can further assume that
\begin{equation*}
  P(\sgn(\Delta E) = 1) \approx P(\sgn(\Delta E) = - 1) \approx 0.5~ .
\end{equation*}

With this, equation (\ref{eq:accrate}) can be written as
\begin{equation*}
  r_A(\epsilon) = \frac 12 \left( 1 +
    \left\langle e^{-|\Delta E|} \right\rangle_{P(|\Delta E| | \epsilon)}\right)
  \approx \frac 12 \left( 2 - \left\langle |\Delta E| \right\rangle \right)
  + \order(\langle |\Delta E|^2 \rangle)\,,
\end{equation*}
where the exponential-function was expanded to first order.

In practice a certain value for $r_A$ like $0.8$ is recommended.
This means for $\Delta E$
\begin{equation}\label{eq:derelation}
 \langle |\Delta E| \rangle \overset != 2 (1 - r_A(\epsilon)) =: \tde
\end{equation}
This is the relation that lies at the core of most epsilon-adjusting
strategies available in \hmcif .
Note that even if a step is accepted for sure because $\Delta E$ is negative,
adjusting $\epsilon$ is still possible since only the absolute value of $\Delta E$ is needed.
This is of great use in cases where nearly every step during burn-in
produces a negative $\Delta E$
(This happens sometimes if the Markov chains start far off the mean value).
For more on how $\epsilon$ is adjusted exactly see section \ref{sec:eps}.

%% file: tex/software_structure.tex
\section{Detailed Software Description}\label{sec:soft}
This section describes all features and functionalities of \hmcif .
\subsection[The HMCSampler Class]{The \hmcsc\ Class}
The \hmcif\ package is optimized for a fast HMC implementation for a \niftye\ class.
At its heart lies the \hmcsc\ class handling the whole sampling process.
It is able to run several Markov chains on different CPUs
using the \python\ \pkg{multiprocessing} module.
The samples are saved to an HDF5-file which is generated every time a new run is initialized
and can be loaded back as needed via the package's own \code{load} function.
During the burn-in phase \hmcsc\ takes care of adjusting the integration
step size $\epsilon$ (see equation (\ref{eq:leapfrog})) such that a user
defined acceptance rate is reached,
as well as setting and possibly reevaluating the mass matrix $M$.
After a run has finished the mean of the samples is calculated.

Of course in practice one may want to fine-tune some of the specific features
and parameters implemented in \hmcif .
This section is dedicated to introduce and explain those.

\subsubsection{Instantiating}

An instance of \hmcsc\ is created with the following arguments of which only the first is mandatory:
\begin{description}
 \item[\code{potential} :]
  \niftye\ \\
  The HMC potential $\Psi(x)$.
  Also defines the domain on which the sampling takes place through its position attribute.
 \item[\code{sample\_transform} :]
  \code{func}, optional\\
  In some cases it is preferable to sample a field not in its position space,
  but in another domain, such as in its harmonic space representation,
  or maybe even in a domain where there is no linear transformation
  to the position space.
  To ensure correct calculation of expectation values such as the mean or the variance
  the samples are transformed by \code{sample\_transform} before being saved to disk.
  The \code{sample\_transform} function has to have exactly one argument which
  is a \code{Field} or \code{MultiField} similar to the \code{position}
  attribute of the \nifty\ \eon\ given as the \code{potential} argument.
  The \code{sample\_transform} function has to return a \code{Field} or
  \code{MultiField}.
  There are, however, no further restrictions on the exact structure of this
  \code{Field} or \code{MultiField}.
 \item[\code{num\_processes} :]
  \code{int}\\
  Number of cores involved in the sampling process.
  This is equal to the number of individual Markov chains started when the instance method \code{run} is called.\\
  Default: 1
 \item[\code{sampler\_dir\_path} :]
  \code{str}\\
  A path where the HDF5-file containing the samples is going to go.\\
  Default: a new folder called ``samples'' in the `\code{\_\_main\_\_}' script's directory
\end{description}

\subsubsection{Running the Sampling Process}
In principle the sampling process can be started immediately
after creating an instance of \hmcsc\ by calling the \code{run()} method
with the following arguments of which again only the first is mandatory.

\begin{description}
 \item[\code{num\_samples} :] 
  \code{int}\\
  Number of samples to be drawn per chain after burn in. 
 \item[\code{max\_burn\_in} :] 
  \code{int}, optional\\
  Maximum number of steps for the chain to converge before it is forced into sampling mode. 
  If no value is stated, forced transition is not going to happen.
  In this case the chain will only start the actual sampling process if it has converged.
 \item[\code{convergence\_tolerance} :]
  \code{float}, optional\\
  If the convergence measure for the sampling process falls below this value,
  the chain is assumed to have converged
  and starts with the actual sampling process.
  If no value is stated the \code{tolerance} property of
  the \hmcsc\ property \code{convergence} is used.
  The default value for said property is 1.
  For more on this see section \ref{sec:conv}.
 \item[\code{target\_acceptance\_rate} :]
  \code{float}, optional\\
  Value between 0. and 1., stating what ratio of accepted / rejected samples is preferred.
  The integration step size is adjusted during burn-in to approximately match this ratio.\\
  If not stated the corresponding property of the \code{epsilon} property is
  used (for which the default value is 0.8).
 \item[\code{order} :]
  \code{int}\\
  The order of the symplectic integrator.
  The default value corresponds to a simple leapfrog integration.
  Default: 2
 \item[\code{mass} :]
  \nifty\ \code{EndomorphicOperator}, optional\\
  HMC mass matrix used during sampling (or until it is reevaluated).
  For more on the mass matrix see section \ref{sec:mass}.
  If no mass is given, an identity matrix is used (at least as initial guess).
 \item[\code{x\_initial} :]
  \niftyf\ or \pylist\ of \niftyf s, optional\\
  Starting point(s) for the HMC sampler. 
  If more than one Markov chain needs to be initialized
  they get their respective initial positions by iterating through the list.
  The list does not have to have the same length as the number of chains.
  If there are more chains than elements in the \pylist ,
  some starting positions are reused for the additional chains.
  If only a \code{Field} is given, all chains get the same initial position.
  If no initial field is passed,
  a random sample is drawn from a Gaussian distribution centered at the position of the \code{Energy} instance
  given to the constructor of \hmcsc\ with the \code{Energy}'s metric as covariance.
\end{description}

\subsubsection{Getting the Results of an HMC-Run}
After a run has finished, the sampled mean as a \niftyf\ or \code{MultiField} is
accessible via the instance's property `\code{mean}'.

\begin{lstlisting}[language=iPython]
    |\iin| hmc_sampler = HMCSampler(nifty_energy, num_processes=5)
    |\iin| hmc_sampler.run(200)
    |\iin| hmc_sampler.mean
    |\out| nifty4.Field instance
           - domain  = DomainTuple, len: 1
           RGSpace(shape=(256, 256), distances=(0.5, 0.5),
                   harmonic=True)
           - val     = array([[ 0.63, -0.16,  ...,  1.04, -0.64],
                              ..., 
                              [ 0.03 ,  0.02,  ...,  1.22,  0.21  ]])
\end{lstlisting}

Accessing the values of the samples is possible via calling the \code{samples}
property.
It consists of a $2+n$ dimensional \np\ \code{ndarray} where the first
dimension represents the different Markov Chains,
the second dimension represents the individual samples
and the other $n$ dimensions represent the value of the sampled \nifty\ \code{Field}s.
If the sampled objects were \code{MultiField}s an dictionary with \np\
\code{ndarray}s is returned.
Remember though that calling this will load all samples into memory
which might crash the process if not enough memory is available.

\ContinueLineNumber
\begin{lstlisting}[language=iPython]
    |\iin| all_samples = hmc_sampler.samples
    |\iin| all_samples.shape
    |\out| (5, 200, 256, 256)
\end{lstlisting}

\subsubsection[Attributes and Properties of HMCSampler]{Attributes and Properties of \hmcsc}

The \hmcsc\ has a number of other properties and attributes which are mostly used for fine-tuning the sampling
process.
These are:
\begin{description}
  \item[\code{potential} :]
    \nifty\ \eon , read-only\\
    The potential $\Psi$ for the HMC sampler.
  \item[\code{sampler\_dir} :]
    \code{str}\\
    Setting or getting the path to the directory where the sample-files are stored.
    Corresponds to the parameter of the same name passed in the constructor of \hmcsc\ class. 
  \item[\code{save\_burn\_in\_samples} :]
    \code{bool}\\
    Whether or not to save the samples generated during burn-in phase to disk.
    Be aware of the fact that if set to \code{True} (default value) together with a high or
    non-existent \code{max_burn_in} parameter (in the constructor of \hmcsc\
    class)
    could fill your hard drive.\\
    Default: True
  \item[\code{burn\_in\_samples} :]
    \np\ \code{ndarray} or \code{dict}(\code{str -> ndarray}), read-only\\
    The same as the \code{samples} property but with the samples generated
    during burn-in phase.
  \item[\code{var} :]
    \nifty\ \code{Field} or {MultiField}, read-only\\
    The variance of the samples (after \code{sample\_transform}).
  \item[\code{convergence} :]
    \hmcif\ \convc\\
    For choosing how to calculate the convergence measure (see section \ref{sec:conv}).\\
    Default: \code{HansonConvergence if num_processes == 1 else GelmanRubinConvergence} 
  \item[\code{epsilon} :]
    \hmcif\ \epsc\\
    For choosing how to adjust the integration step size parameter
    during burn-in (see section \ref{sec:eps}).\\
    Default: \code{EpsilonPowerLawDivergence}
  \item[\code{mass} :]
    \hmcif\ \massc\\
    For choosing how to handle the HMC mass during sampling. 
    For more on this see section \ref{sec:mass}.
  \item[\code{display} :]
    \hmcif\ \dispc\\
    For choosing how to display the progress of the sampling process (see section \ref{sec:disp})\\
    Default: \code{LineByLineDisplay}
  \item[\code{n\_limits} :]
    \pylist\ of \integer s\\
    To avoid periodic trajectories the number of
    leapfrog integration steps is randomized.
    \code{n\_limits} defines the range from which the number of
    integration steps is drawn uniformly.\\
    Default: \code{[60, 70]}
\end{description}

\subsection[Convergence]{\convc}\label{sec:conv}
The \convc\ class handles everything related to the convergence of the Markov chain(s).
In principle a chain in \hmcif\ has converged if a \emph{convergence measure}
calculated for each degree of freedom in the sampled \niftyf\ or \code{MultiField}
drops below a given \emph{tolerance}.

Additionally \hmcif\ implements intermediate steps of convergence
via so-called convergence \emph{levels}.
For the time being their main purpose is to define a time
where the HMC mass is reevaluated during burn-in
(See also section \ref{sec:mass}).
A chain is said to have converged with respect to its current
convergence level, if
\begin{equation}\label{eq:converged}
  \max(\text{measure}) < \text{tolerance} \cdot 10^\text{level}
\end{equation}
In other words:
The level is the number of digits the decimal separator
of the tolerance is shifted to the left.

The idea behind this is to decrease the level by one
each time an intermediate convergence is reached,
while at that point recalculating the mass matrix.

If the level drops to zero, equation (\ref{eq:converged})
simplifies to $\max(\text{measure}) < \text{tolerance}$
and the next time this requirement is met,
the Markov chain has finished the burn-in phase.

It remains to explain how the convergence measure is calculated.
There are several different approaches implemented in \hmcif\
as child classes of the \convc\ class.
Choosing one of them is done by setting
the \code{convergence} property of the \hmcsc\ class with
one of the \convc 's child classes, e.g.:
\begin{Code}
hmc_sampler.convergence = GelmanRubinConvergence
\end{Code}
For now there are four different possibilities:
\begin{description}
  \item[\code{MeanConvergence}]
    This rather simple way of calculating the convergence needs at least two Markov chains.
    It compares the mean of the samples from all chains (\code{total\_mean}) to the mean of each individual chain (\code{chain\_means}).
    The measure is defined as \code{abs(chain\_mean / total\_mean - 1.)}
    such that it fulfills the non-negativity and the identity of indiscernibles criteria for metrics.
    It proves to be rather unstable if e.g. the total mean is close to zero.
  \item[\code{VarConvergence}]
    Very similar to \code{MeanConvergence} only with the variances of individual chains and all chains.
    Measure is equal to \code{abs(chain\_var / total\_var - 1.)}.
  \item[\code{HansonConvergence}]
    So far the only convergence measure which can be used even if there is only one chain.
    It follows \citet{hanson2001}
    (Again: the measure is the absolute value of the ratio minus one).
  \item[\code{GelmanRubinConvergence}]
    An implementation of the among MCMC folks very popular
    Gelman and Rubin convergence criteria \citet{gelman1992}
   (Again: the measure is the absolute value of the ratio minus one).
\end{description}

\subsubsection[Attributes and Properties of Convergence]{Attributes and Properties of \convc}
Regardless of which \convc\ child class has been used additional features can be set
via its class properties, e.g. the \code{locality} property which defines
the number of recent samples considered when calculating the \code{convergence}
(see below for details):
\begin{Code}
hmc_sampler.convergence = GelmanRubinConvergence
hmc_sampler.convergence.locality = 200
\end{Code}
For the common user the following properties are the most important ones:
\begin{description}
  \item[\code{locality} :]
   \integer\\
   The number of recent samples to be considered in calculating the convergence
   measure.
   On the one hand this is a form of `forgetting' very old parts of the chain's
   trajectory which do not represent the current state of convergence.
   On the other hand this is necessary because of memory issues
   i.e. if the burn-in phase takes very long the memory would blow up
   since every sample ever created has to be available to calculate the
   measure.\\
   Default: 250
  \item[\code{tolerance} :]
   \float \\
   Equivalent to the \code{convergence\_tolerance} parameter
   of the \hmcsc 's \code{run} method.    
   In fact, setting this property as described above has only an effect
   if the (optional) \code{convergence\_tolerance} parameter
   is not passed to the \code{run} method.\\
   In practice the latter approach might be slightly more convenient.
   If the maximum value of a chain's convergence measure is below this value
   the chain is said to have converged and transitions from the burn-in phase
   into the sampling phase.
   See also: \code{converged} (below)\\
   Default: 1.
\end{description}
The following additional properties of \convc\ are mostly only important for
\hmcsc\ itself and not of relevance for the common user:
\begin{description}
 \item[\code{converged} :] \nparray\ of \bool\ (1 dim)\\
   Contains the information of whether the individual chains have converged
   with respect to the following law:\\
   \code{converged = measure\_max < tolerance * 10**level}\\
 \item[\code{measure} :] \nparray\ ($1 + n$ dim)\\
   Represents the value of the measure
   (calculated dependent on which child class of the \convc\ class has been used)
   for each element of the degrees of freedom in the sampled \niftyf .
   The first dimension represents the individual chains.
 \item[\code{measure\_max} :] \nparray\ ($1$ dim)\\
   The highest value of the \convc\ class property `\code{measure}' per chain.
 \item[\code{level} :] \nparray\ of \integer\ (1 dim)\\
   See class property \code{converged}.
   The idea is that after a Markov chain has converged
   with respect to its current level the level is decreased by one.
   There are \convc\ class methods \code{dec\_level} and \code{inc\_level}
   for decreasing and increasing the level by 1, respectively.
   For more details on these methods see below.\\
   Setting this property is also possible with a simple \integer\
   which sets the whole \nparray\ to that value.
 \item[\code{quota} :] \nparray\ of \float (1 dim)\\
   The ratio of elements in the sampler's position \niftyf\ which have converged
   with respect to \code{tolerance} and \code{level}
   (i.e. the 'intermediate' convergence)
\end{description}

\subsubsection[Additional Methods of Convergence]{Additional Methods of \convc}
Internally the convergence levels are decreased and increased by calling
\begin{description}
\item[\code{dec\_level(chain\_identifier=None)}] \hfill \\
  Decreases the convergence level of \code{chain\_identifier} (\integer )
  by one. If \code{chain\_identifier} is \code{None}
  the level of all chains is decreased by one.
  Either way if the level of a chain is already zero it is left unchanged.
\item[\code{inc\_level(chain\_identifier=None)}] \hfill \\
  Increases the convergence level of \code{chain\_identifier} (\integer )
  by one. If \code{chain\_identifier} is \code{None}
  the level of all chains is increased by one.
\end{description}
The convergence level is set under the hood
dependent on specific properties of the \massc\ class
in the beginning of the \hmcsc 's \code{run} method.

\subsection{Epsilon}\label{sec:eps}
The $\epsilon$ parameter defines the leapfrog integration step size
(equation (\ref{eq:leapfrog})).
In principle the bigger it is the bigger is the integration error $\Delta E$
and thereby the smaller the acceptance rate.
To achieve an approximate acceptance rate
defined via the \code{target\_acceptance\_rate} parameter of
\hmcsc 's \code{run} method, $\epsilon$ has to be adjusted during burn-in.
Every adjusting-strategy relies on thoughts presented in appendix
\ref{sec:epsadj} connecting the \code{target\_acceptance\_rate} to the
integration error $\Delta E$.
In \hmcif\ the \epsc\ class, much like the \convc\ class, is just a base class
and much more interesting for the common user are its child classes defining
exactly how $\epsilon$ is adjusted.

The class also keeps track of how much $\epsilon$ has changed in recent steps
and how close the mean value of recent integration errors
$\langle \Delta E \rangle$ is to $\tde$.
If $\epsilon$ has not changed very much and
$\langle \Delta E \rangle \approx \tde$, \epsc\ is said to have converged.

If \epsc\ has converged its value is locked.

\subsubsection{Available Adjusting Strategies}
\begin{description}
\item[\code{EpsilonConst}] 
  $\epsilon$ stays constant throughout the whole sampling process.
  The value can be set by setting its \code{val} attribute:
  \begin{Code}
hmc_sampler.epsilon = EpsilonConst
hmc_sampler.epsilon.val = 0.005
  \end{Code}
\item[\code{EpsilonSimple}] 
  $\epsilon$ gets reduced or increased if $\Delta E$ is bigger or smaller
  than $\tde$ respectively independent of the absolute value of $\Delta E$.

  In practice, \code{EpsilonSimple} has an attribute \code{change_range}
  (\float\ between 0 and 1, Default: 0.1), which can be set via:
  \begin{Code}
hmc_sampler.epsilon = EpsilonSimple
hmc_sampler.epsilon.change_range = 0.2
  \end{Code}
  This attribute is only available in \code{EpsilonSimple}.
  Given the \code{change_range} the current value of $\epsilon$
  is multiplied by a factor drawn from
  a uniform distribution $\uni ([a, b])$ where
  \[
    [a, b] = \left\{
        \begin{array}{ll}
          {[1 - \text{\code{change\_range}}, 1]} & \text{if } \Delta E > \tde\\
          {[1, 1 + \text{\code{change\_range}}]} & \text{if } \Delta E < \tde
        \end{array}
        \right.
  \]
  The randomness is necessary to prevent recurrent behavior
  if the integration error $\Delta E$ is very sensitive to $\epsilon$.
\item[\code{EpsilonPowerLaw}] 
  $\epsilon$ gets adjusted just like \code{EpsilonSimple} but the
  \code{change\_range} is now defined by the relative difference between
  $\Delta E$ and $\tde$
  (\code{EpsilonPowerLaw} has no attribute \code{change_range}!).\\
  Given this class's attribute \code{power} (positive \integer , Default: 5),
  set via
  \begin{Code}
hmc_sampler.epsilon = EpsilonPowerLaw
hmc_sampler.epsilon.power = 4
  \end{Code} 
  the \code{change\_range} in \code{EpsilonSimple} is defined as:
  \begin{equation}\label{eq:epspower}
    \text{\code{change\_range}} = \left| \frac{\Delta E - \tde}{\Delta E + \tde} \right|^{\text{power}}
  \end{equation}
\item[\code{EpsilonPowerLawDivergence}]
  In practice working with Poissonian or log-normal distributions on high
  dimensional spaces the integration error $\Delta E$ proved to be
  very sensitive to small changes in $\epsilon$. 
  With this class $\epsilon$ is adjusted just like \code{EpsilonPowerLaw}
  with the difference,
  that in case of a divergent $\Delta E$ (e.g. during integration an overflow
  occurs) the \code{change\_range} becomes more sensitive.

  A \code{divergence\_counter} keeps track of the number of times a
  divergent behavior was detected and the \code{change\_range} defined
  in equation (\ref{eq:epspower}) gets a prefactor $2^{-\text{\code{divergence\_counter}}}$
\item[\code{EpsilonExponential}] 
  In this case a simple connection between $\Delta E$ and $\epsilon$ is assumed:
  \begin{equation}\label{eq:epsexp}
    |\Delta E|(\epsilon) = a \cdot \epsilon^b
  \end{equation}
  where $a > 0$ and $b > 0$ are fitting parameters.
  This assumption is motivated by the fact that $\Delta E$ tends to zero
  if $\epsilon$ does and diverges for large $\epsilon$.
  Former `measurements', i.e. sampling steps of
  $\Delta E$ given $\epsilon$ are used to calculate $a$ and $b$.
  This approach asks for a rather large value for the \code{locality}
  property (see below).
  $\epsilon$ is adjusted by rearranging equation (\ref{eq:epsexp})
  such that:
  \begin{equation}
    \epsilon_{\text{new}}= \left( \frac 1a \left|\tde \right| \right)^{\frac 1b}
  \end{equation}
\item[\code{EpsilonOneOverX}] 
  In this case another connection between $\Delta E$ and $\epsilon$ is assumed:
  \begin{equation}\label{eq:epsx}
    |\Delta E|(\epsilon) = \frac{a}{(\epsilon_0 - \epsilon)^b} + \text{const}
  \end{equation}
  where $a > 0$ and $b > 1$ are again fitting parameters and
  $\text{const}$ is such that $|\Delta E|(\epsilon=0) = 0$.
  The idea behind this relation is an updated version of \code{EpsilonExponential}
  where there is a finite $\epsilon_0$ for which $\Delta E$ diverges already.
  $a$ and $b$ are again fitted with former $\Delta E$s given $\epsilon$,
  whereas $\epsilon_0$ is set to the current value of $\epsilon$
  every time a divergent behavior is detected.
  $\epsilon$ gets adjusted by rearranging equation (\ref{eq:epsx}), such that
  \begin{equation}
    \epsilon_{\text{new}} = \epsilon_0 \left( 1 - \left( 1 + \frac{\epsilon_0^b \tde}{a} \right)^{\frac 1b} \right)
  \end{equation}
\end{description}

\subsubsection[Attributes and Properties of Epsilon]{Attributes and Properties of \epsc}
Regardless of which \epsc\ class has been used, additional features can be set
via its class properties, e.g. the \code{locality} property which defines
the scope of fitting points for classes like \code{EpsilonExponential} and
for calculating the convergence measure
(see below for details):
\begin{Code}
hmc_sampler.epsilon = EpsilonPowerLawDivergence
hmc_sampler.epsilon.locality = 20
\end{Code}
For the common user the following properties and attributes
are the most important ones:
\begin{description}
  \item[\code{val} :]
    \float \\
    The (current) value of $\epsilon$.
    It is possible to use this property to set a good initial guess
    (although most of the time unnecessary). \\
    Default: 0.005
  \item[\code{locality} :]
   \integer\\
   The number of recent samples to be considered in calculating the convergence
   measure of \code{epsilon}.\\
   Default: 50
  \item[\code{target\_acceptance\_rate} :]
   \float \\
   Equivalent to the \code{target\_acceptance\_rate} parameter
   of the \hmcsc 's \code{run} method.
   In fact, setting this property only has an effect
   if the (optional) \code{target\_acceptance\_rate} parameter
   is not passed to the \code{run} method.\\
   Default: 0.8
  \item[\code{convergence\_tolerance} :]
    \float \\
    Essentially the same thing as the \code{tolerance} property of
    \code{Convergence} (section \ref{sec:conv}) but for \code{epsilon}
    A value $> 1$ is unreasonable because of how the convergence measure
    for \code{epsilon} is calculated (see below)\\
    Default: 0.5
  \item[\code{divergence\_threshold} :]
    \float \\
    The value of $\Delta E$ for which the integration is said to have diverged.\\
    Default: $1E50$
  \item[\code{epsilon\_limits} :]
    \pylist\ of \float \\
    Minimum and maximum value for \epsc .
    If the adjusting algorithm proposes a value `out of range'
    the value gets coerced.\\
    Default: \code{[1.E-40, 10]}
\end{description}
Under the hood \hmcsc\ uses the following additional properties of \epsc :
\begin{description}
 \item[\code{converged} :] \bool , read-only\\
   Whether or not \code{epsilon} has converged or not,
   i.e. whether the \code{convergence\_measure} is smaller than the
   \code{convergence\_tolerance}
 \item[\code{measure} :] \code{list} of \float , read-only\\
   The convergence measure for \code{epsilon} contains two values.
   The first is the relative variance of the value $\epsilon$,
   the second represents how close the mean value of $\Delta E$
   is to $\tde$.
   If both are smaller than \code{convergence_tolerance},
   \code{epsilon} has converged.
\end{description}

\subsection{Mass}\label{sec:mass}
Finding an appropriate mass matrix is the most challenging
task for a good HMC sampler.
\hmcif\ provides the user with the standard evaluation procedure
introduced in equation (\ref{eq:masseval}) as well as the
possibility to define a problem specific mass matrix.

(Re-)evaluation is done by drawing samples $\{x^{(i)}\}$ from the curvature
at the current position of the \nifty\ \eon\
given as \code{potential} parameter in the constructor.
To keep the complexity of the problem bearable
only the diagonal of the mass matrix in equation (\ref{eq:masseval})
is calculated and used:
\begin{equation}\label{eq:massevaldiag}
  M_{jk} = \delta_{jk} \frac 1{N-1} \left( \sum_{i=1}^N {x_j^{(i)}}^2 \right)^{-1}
\end{equation}
This also removes the problem that for a non-degenerate mass matrix in
$n$ dimensions at least $n$ independent samples are required.
For typical applications in \nifty\ $n=10^6 .. 10^8$ easily.

The idea behind the \hmcif\ \massc\ class is to easily define
a strategy of handling the mass matrix of an HMC process.
This includes reevaluating the mass matrix several times
during burn-in phase and defining an initial mass matrix.
As default the identity is used as mass matrix but an initial
mass matrix can be evaluated without reaching any level of convergence.
By default there is one such initial mass evaluation and no reevaluations.

\begin{description}
\item[\code{get\_initial\_mass} :]
  \bool \\
  Whether or not to evaluate an initial mass.
  Setting this to \code{True}/\code{False} will increase/decrease
  \code{reevaluations} by 1 respectively.\\
  Default: \code{True}
\item[\code{reevaluations} :]
  \nparray\ of \integer \\
  The number of reevaluations (including the initial one if set)
  for each chain.
  Reevaluation takes place if the chain has converged with respect to
  its current convergence level introduced in section \ref{sec:conv}.
  Setting the number of reevaluations is also possible (and recommended) with a
  simple \integer\   which sets all chains to that \integer .\\
  Default: \code{numpy.ones(num_chains)}
  (i.e. one reevaluation for every chain)
\item[\code{operator} :]
  \nifty\ \code{EndomorphicOperator}\\
  The actual mass operator.
  If there is a problem specific mass operator it can be set with this property
  before initiating the run.
  If \code{get_initial_mass} is \code{True},
  setting \code{operator} will set \code{get_initial_mass} to \code{False}
  and decrease \code{reevaluations} by 1.\\
  Default: Identity (as \nifty\ \code{DiagonalOperator}) 
\item[\code{num\_samples} :]  
  \integer \\
  Defines the number of samples drawn from the curvature
  when reevaluating the mass matrix.\\
  Default: 50
\item[\code{shared} :]
  \bool \\
  If \code{True} reevaluating a mass matrix is done at the mean
  of all individual chain positions.
  All chains have to meet the conditions for evaluation mentioned above.
  Afterwards each chain gets the same mass matrix.
  If \code{False} each chain reevaluates its mass matrix individually
  if the chain meets the conditions for evaluation.\\
  Default: \code{False}
\end{description}

\subsection{Display}\label{sec:disp}
Naturally HMC sampling can take quite a while.
Disadvantageous settings of parameters might lead to a malfunctioning
sampling process.
To be able to discover a pathological run can save hours or even days.
For this reason \hmcif\ offers a number of display modes for diagnostic purposes.
On the other hand displaying indicators of tens of parallel running
chains might be overwhelming.
All available display options have in common that they display more or less
information based on a \emph{logging level} just like the logging levels in
the \pkg{logging} package \citep{pylog} of the \python\ standard library.
It is in fact possible to use the appropriate module variables from the
\pkg{logging} package.
These are:
\begin{description}
\item[\code{CRITICAL}] (numerical value: \code{50})
\item[\code{ERROR}] (numerical value: \code{40})
\item[\code{WARNING}] (numerical value: \code{30})
\item[\code{INFO}] (numerical value: \code{20})
\item[\code{DEBUG}] (numerical value: \code{10})
\item[\code{NOSET}] (numerical value: \code{0})
\end{description}
The logging level can be set using the \code{display}s method \code{setLevel} e.g.,
\begin{Code}
from logging import INFO
hmc_sampler.display.setLevel(INFO)
\end{Code} 

\begin{figure}
\centering 
\begin{subfigure}{0.9\textwidth}
\includegraphics[width=0.9\linewidth]{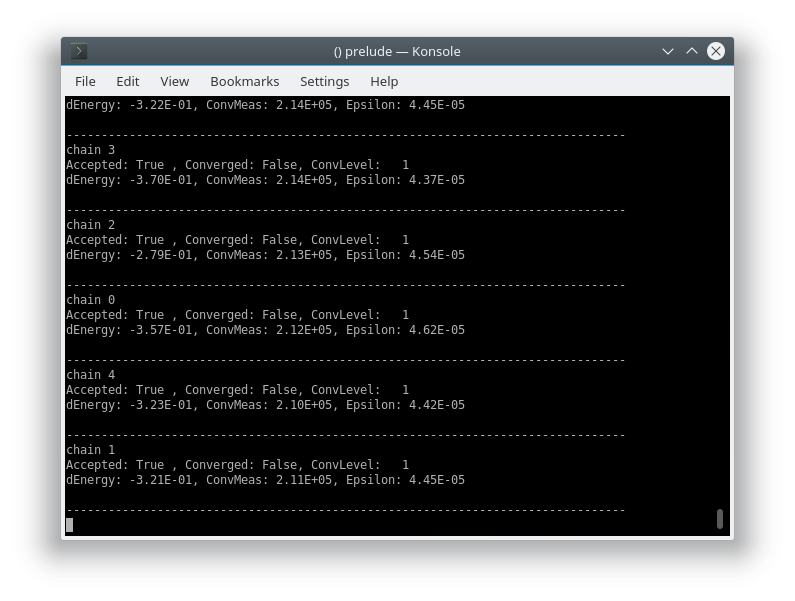}
\caption{\code{LineByLineDisplay} with \code{level = DEBUG} during burn in}
\label{fig:lbld}
\end{subfigure}
 

 
\begin{subfigure}{0.9\textwidth}
\includegraphics[width=0.9\linewidth]{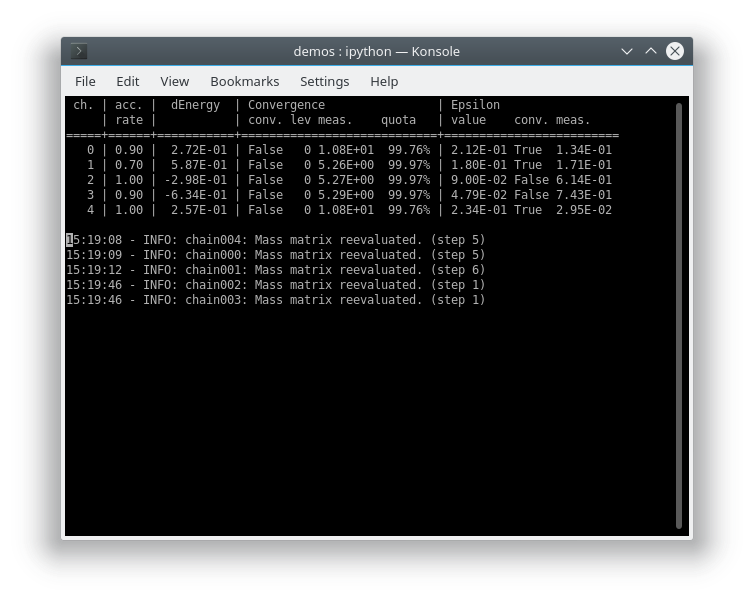} 
\caption{\code{TableDisplay} with \code{level = DEBUG} during burn in}
\label{fig:td}
\end{subfigure}
 
\caption{}
\label{fig:displays}
\end{figure}

Similar to the \epsc\ and \convc\ classes there are several \dispc\
classes which define the three displaying modes.
\begin{description}
\item[\code{Display} :]
  Displays nothing at all.
  Serves as base class for the two other display classes.
\item[\code{LineByLineDisplay} :]
  If the level of the \hmcif\ logger is set to \code{INFO} or below 
  certain indicators are printed at every sampling step
  as depicted in figure \ref{fig:lbld}.
  Otherwise only warnings and errors are printed.
\item[\code{TableDisplay} :]
  The most advanced version of displaying indicators.
  A table is generated and dynamically updated, containing information
  for each chain as depicted in figure \ref{fig:td}.
  This class relies heavily on the \pkg{curses} \python\ package and
  therefore alters the terminal behavior during sampling.
\end{description}

The columns in \code{TableDisplay} display the following parameters:\\
\begin{tabularx}{\textwidth}{lX}
  \code{ch}        & The chain number or `identifier'.\\
  \code{acc.\ rate} & The acceptance rate for each chain.
  During burn in this is only the acceptance rate of the last 10 samples
  since this highly depends on the value of \epsc .
  After burn in \code{acc.\ rate} displays the overall acceptance rate.\\
  \code{dEnergy} & The most recent value of the integration error
  $\Delta E$.\\
  \code{Convergence} & \\
  \utab\code{conv.} & Whether the chain has converged with respect to the current
  convergence level.\\
  \utab\code{lev} & The current convergence level.\\
  \utab\code{meas} & The maximum value of the convergence level of the chain.\\
  \utab\code{quota} & The percentage of points in the sampled \niftyf\ which have
  converged with respect to the current convergence level.\\
  \code{Epsilon} & \\
  \utab\code{value} & The current value of \epsc . \\
  \utab\code{conv.} & Whether or not \epsc\ has converged with respect to its measure.\\
  \utab\code{meas.} & The maximum value of the \epsc\ convergence measure.
\end{tabularx}

\subsubsection[Additional Methods of Display]{Additional Methods of \dispc}
Regardless of which \dispc\ class has been used the quantity of displayed
information can be changed via the \code{setLevel} method.
\begin{description}
  \item[\code{setLevel(level)}] \hfill \\
   This is equivalent to the \code{setLevel} function in the \python\ package
   \pkg{logging}.
   Default: \code{logging.INFO}
\end{description}

\subsection{Additional Functions}\label{sec:addfunc}
\hmcif\ provides a number of additional functions targeted at
easing the handling of the HDF5-files generated during sampling.
These files have rather cryptic
default names of the form `runYYYY-MM-DD\_hh-mm-ss.h5',
where Y, M, D, h, m and s represent year, month, day, hour, minute and second
digits at the time of the initialization of that run, respectively.

There are three functions handling data stored in these files:
\code{load}, \code{load\_mean}, \code{load\_var}

\subsubsection[The load Function]{The \code{load} Function}
The \code{load} function takes the following arguments
of which only the first is mandatory.
Keep in mind that for high dimensional problems the amount
of data loaded to memory can easily be several GiBytes.
\begin{description}
\item[\code{path} :] 
  \str \\
  Either the path to the HDF5 file
  or a path to the directory where the HDF5 file(s) are stored.
  If \code(path) is a directory, the latest `run' file
  (with respect to the file name) is loaded.
\item[\code{attr\_type} :]
  \str \\
  The `attribute' to load.
  For files generated with the \hmcsc\ class possible values are:
  \code{'burn_in'} and \code{'samples'}.
  For \code{'burn_in'} the samples generated during burn in are loaded
  (of course, this is only possible if the \hmcsc\ class attribute
  \code{save_burn_in_samples} was set to \code{True}).
  For \code{'samples'} the samples generated after burn in are loaded.
  Default: \code{'samples'}
\item[\code{start} :]
  \integer \\
  Only loads the samples from step \code{start} onward.\\
  Default: 0
\item[\code{stop} :]
  \integer , optional\\
  Only loads the samples up to step \code{stop}.
  Loads until the end if no value is passed
\item[\code{step} :]
  \integer \\
  Only loads every $n$th sample, where $n$ is given by \code{step}. \\
  Default: 1
\end{description}
In all cases the function returns a \nparray\ if \code{Field}s were sampled and
\python dictionaries if \code{MultiField}s were sampled.
In the latter case the respective \nparray\ for each individual field can be
obtained by using the key used in the \code{MultiField}.
In all cases the array has $2+n$ dimensions,
where again, the first and second dimension represent chains and steps
respectively.
If not all chains have the same number of samples
(e.g. \code{attr_type = 'burn_in'}, every chain needs a different number of
steps to reach convergence)
shorter chains are filled with \code{numpy.nan}s in the output array
to match the size of the longest chain.

\subsubsection[The load_mean Function]{The \code{load\_mean} Function}
The \code{load\_mean} function calculates the mean value based
on the samples saved to a HDF5 file.
\begin{description}
\item[\code{path} :] 
  \str \\
  Either the path to the HDF5 file
  or a path to the directory where the HDF5 file(s) are stored.
  If \code(path) is a directory, the latest `run' file
  (with respect to the file name) is loaded.
\item[\code{domain} :]
  \niftyd\ or \nifty\ \code{MultiDomain} , optional\\
  If \code{domain} is given the function output is a \niftyf\ or a \nifty\
  \code{MultiField}
  with the respective domain and the calculated mean as value.
\end{description}
The function returns the mean value either as \nparray\, a \python\ dictionary
of \nparray s, a \niftyf\ or a \nifty\ \code{MultiField} dependent on what was
sampled originally and whether the \code{domain} argument is given.

\subsubsection[The load_var Function]{The \code{load\_var} Function}
The \code{load\_var} function calculates the variance value based
on the samples saved to an HDF5 file.
\begin{description}
\item[\code{path} :] 
  \str \\
  Either the path to the HDF5 file
  or a path to the directory where the HDF5 file(s) are stored.
  If \code(path) is a directory, the latest 'run' file
  (with respect to the file name) is loaded.
\item[\code{domain} :]
  \niftyd , optional\\
  If \code{domain} is given the function output is a \niftyf\
  with domain \code{domain} and the calculated variance as value.
\end{description}
The function returns the variance either as \nparray\, a \python\ dictionary
of \nparray s, a \niftyf\ or a \nifty\ \code{MultiField} dependent on what was
sampled originally and whether the \code{domain} argument is given.

\subsection{Tools}\label{sec:tools}
\pkg{Tools} is a sub-module of \hmcif\ which provides two handy functions
to evaluate an already finished sampling process.
The one, \code{show_trajectories}, provides a GUI to quickly visualize
the individual Markov chains, the other, \code{get_autocorrelation},
calculates the autocorrelation of the chains.

\subsubsection[The show_trajectories Function]{The \code{show\_trajectories} Function}
A simple GUI for visualizing Markov chains based on one- or
two-dimensional problems.
The GUI is divided into two graphs as displayed in figure \ref{fig:gui01}.
The left one represents the underlying problem space
i.e. its geometry with either the `mean' value or a similar user-defined field.
The right graph displays the trajectory of the different chains
for a selected pixel in the left graph.
A pixel can be selected by either entering its coordinates in the top row
and clicking on show, or by just clicking on it in the left reference
picture.

The function takes the following parameters of which only the first
is mandatory:
\begin{description}
\item[\code{path} :] 
  \str \\
  Either the path to an HDF5 file
  or a path to the directory where the HDF5 file(s) are stored.
  If \code(path) is a directory, the latest `run' file
  (with respect to the file name) in said directory is loaded.
\item[\code{reference\_field} :]
  \niftyf\ or \code{MultiField}, \np\ \code{ndarray} or \code{dict(str ->
    \code{ndarray})}, optional\\
  The field displayed on the left graph of the GUI.
  If none is given, the mean value of the samples is used as reference.
\item[\code{field\_name} :]
  str, optional \\
  If \code{MultiFields} were used, this argument is mandatory and specifies
  which of the sub-fields is supposed to be displayed. 
\item[\code{solution} :]
  \niftyf\ or \code{MultiField}, \np\ \code{ndarray} or \code{dict(str ->
    \code{ndarray})}, optional\\
  In case a `right' answer is available (e.g. a mock data example)
  it can passed here and is displayed in the trajectories graph as
  horizontal line as additional reference.
\item[\code{start} :]
  \integer \\
  Only loads the samples from step \code{start} onward.\\
  Default: 0
\item[\code{stop} :]
  \integer , optional\\
  Only loads the samples up to step \code{stop}.
  Loads until the end if no value is passed
\item[\code{step} :]
  \integer \\
  Only loads every $n$th sample, where $n$ is given by \code{step}. \\
  Default: 1
\end{description}

\subsubsection[The get_autocorrelation Function]{The \code{get\_autocorrelation} Function}
Calculates the autocorrelation of the samples (after burn in)
for a given $t$, where $t$ is the shift in
\begin{equation}
  \text{auto\_corr}[t] = \sum_i x(i)\bar{x}(i+t)
\end{equation}
The function takes the following two arguments:
\begin{description}
\item[\code{path} :] 
  \str \\
  Either the path to an HDF5 file
  or a path to the directory where the HDF5 file(s) are stored.
  If \code(path) is a directory, the latest 'run' file
  (with respect to the file name) in said directory is loaded.
\item[\code{shift} :]
  \integer \\
  The shift $t$ as described above.\\
  Default: 1
\item[\code{field\_name} :]
  str, optional \\
  If \code{MultiFields} were used, this argument is mandatory and specifies
  which of the sub-fields is supposed to be used. 
\end{description}

The function returns a $1+n$ dimensional \nparray\ 
where the first dimension represents the different chains
and the other $n$ dimensions the dimensionality of the problem.